\documentclass[prb,twocolumn]{revtex4}
\usepackage{graphicx}

\def\lnod{La$_2$NiO$_{4+\delta}$}
\def\lsno{La$_{2-x}$Sr$_x$NiO$_4$}
\def\lsnox{La$_{1.725}$Sr$_{0.275}$NiO$_4$}

\begin{document}

\title{Transmission-electron-microscopy study of charge-stripe order in
\lsnox}
\author{Jianqi Li}
\affiliation{Brookhaven National Laboratory, Upton, NY 11973-5000}
\author{Yimei Zhu}
\affiliation{Brookhaven National Laboratory, Upton, NY 11973-5000}
\author{J. M. Tranquada}
\affiliation{Brookhaven National Laboratory, Upton, NY 11973-5000}
\author{K. Yamada}
\affiliation{Institute for Chemical Research, Kyoto University, Gokashou,
Uji, 611-0011 Kyoto, Japan}
\author{D. J. Buttrey}
\affiliation{Department of Chemical Engineering, University of Delaware,
Newark, Delaware 19716}
\date{\today}
\begin{abstract}
We characterize the local structure and correlations of charge
stripes in \lsnox\ using transmission-electron microscopy.  We present
direct evidence that the stripe modulation is indeed one-dimensional
within each NiO$_2$ plane.  Furthermore, we show that individual stripes
tend to be either site-centered or bond-centered, with a bias towards the
former.  The spacing between stripes often fluctuates about the mean,
contributing to a certain degree of frustration of the approximate
body-centered stacking along the $c$-axis.  These results confirm ideas
inferred from previous neutron-diffraction measurements on doped
nickelates, and demonstrate that charge-stripe order is quite different
from the conventional concept of charge-density-wave order.
\end{abstract}
\maketitle

The tendency in hole-doped two-dimensional antiferromagnets for 
charges to order in stripes is an empirically established phenomenon
that is of particular interest for its possible relevance to
superconductivity in the cuprates.\cite{oren00,emer99}  Despite
considerable progress on the experimental front, the theoretical
identification of the dominant effects reponsible for the occurrence of
stripe correlations remains
controversial.\cite{zaan89,kato90,zaan94,low94,cast95,%
naya97,whit98a,hell99,vojt99,khom01,podo02}  For superconducting cuprates,
where the stripes are oriented parallel to the Cu-O bonds, there is
considerable theoretical interest in whether the stripes tend to be
centered on rows of Cu atoms (site-centered stripes) or on rows of O
atoms (bond-centered).\cite{whit98a,vojt99}  In the case of nickelates,
the stripes run diagonally, at 45$^\circ$ to the Ni-O bonds, and
information on the lattice alignment of the stripes has been inferred
from neutron diffraction studies.\cite{woch98}  From measurements on
La$_2$NiO$_{4.133}$ it has been inferred that there is a
temperature-dependent shift from mostly site-centered stripes at low
temperature to all bond-centered stripes above the spin-ordering
temperature.\cite{woch98,tran97b}   It is desirable to test such
inferences with local, real-space images, and to test their generality in
Sr-doped nickelates, which lack the long-range stripe order of
La$_2$NiO$_{4.133}$. 

In this paper we report a transmission-electron-microscopy (TEM) study of
charge stripes in \lsnox, a composition that has been characterized
previously by neutron scattering.\cite{lee01,lee02}  Though not the
composition with the longest correlation length,\cite{yosh00} the
incommensurability of the charge stripe order makes it an interesting
test case.  From the neutron work,\cite{lee02} it is known that the charge
ordering temperature of this sample is $\sim190$~K.  Here we present the
first direct evidence, to our knowledge, that the charge-order modulation
in nickelates is truly one-dimensional.  We also present lattice images
sensitive to the stripes, demonstrating that they frequently exhibit
irregular spacing and that they have a tendency to be either site- or
bond-centered, with a bias towards site-centered in this sample at low
temperature. 

There have been a number of TEM studies of nickelates,
with most\cite{hiro90,oter92,demo92,saya96} focussed on determination of
the interstitial order in \lnod.  We actually spent considerable time
trying to characterize La$_2$NiO$_{4.133}$, and saw some of the
modulations reportedly previously\cite{hiro90,oter92,demo92,saya96};
however, these ordering wave vectors are not all consistent with what has
been observed by neutron diffraction on large crystals.\cite{tran95b}  We
suspect that the lack of consistency between neutron and electron
diffraction measurements is associated with the mobility of the oxygen
interstitials, as suggested by Otero-Diaz and coworkers.\cite{oter92}  In
part due to these difficulties, we turned to \lsno, the system in which
charge ordering was first discovered.\cite{chen93}

Starting with a piece of the \lsnox\ single crystal used in the neutron
study,\cite{lee02} specimens for TEM observations were prepared by
mechanical polishing to a thickness of around 10~$\mu$m, followed by ion
milling.  In addition, we also prepared some thin samples for electron
diffraction experiments simply by crushing the bulk material into fine
fragments, which were then supported by a copper grid coated with a thin
carbon film.  The TEM investigations were performed on a JEOL-3000F
(300~kV) field-emission electron  microscope equipped with an energy
filter.  All of the measurements reported here
were performed with the sample holder cooled by liquid nitrogen, giving an
approximate sample temperature of 86~K.  Images were recorded on either
image plates or a CCD (charge-coupled device) camera. Although the average
crystal structure is tetragonal (space group $I4/mmm$), it is convenient
to index the Bragg reflections with an
$F4/mmm$ unit cell, such that $a=b=5.4$~\AA$=2\sqrt{2}d_{\rm Ni-O}$, with
$c=12.6$~\AA.

\begin{figure}[t]
\centerline{\includegraphics[width=3.2in]{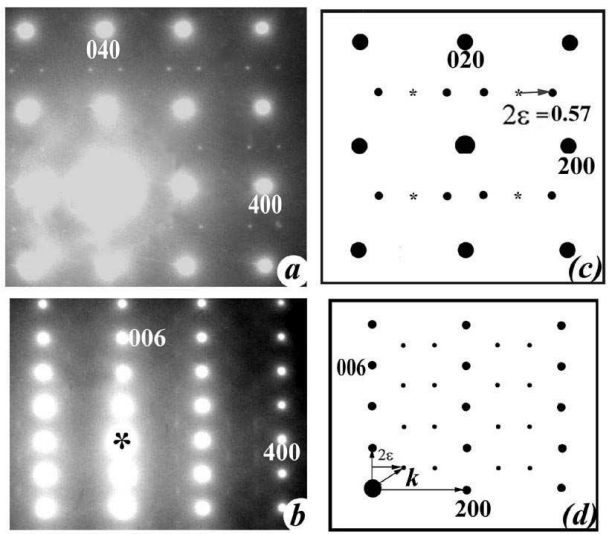}}
\medskip
\caption{Electron diffraction patterns taken along (a) [001] and (b)
[010] zone-axes, showing  the presence of superlattice spots at 86~K.
Schematic illustrations identify the satellite spots in the  reciprocal
lattice planes (c) $a^*$-$b^*$ and (d) $a^*$-$c^*$. }
\label{fg:1}
\end{figure}

\begin{figure}[t]
\centerline{\includegraphics[width=3.4in]{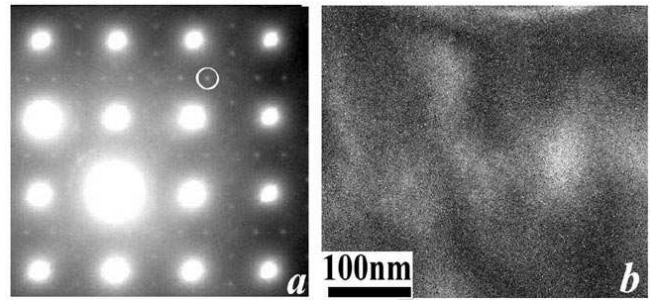}}
\medskip
\caption{(a) Electron diffraction pattern taken from an area showing
$90^\circ$-twinning of superlattice  modulations. (b) Dark-field image
produced by a satellite spot, circled in (a), at the temperature of 86~K,
illustrating the complex charge-ordered domains in the $a$-$b$ plane.}
\label{fg:2}
\end{figure}

Figures~\ref{fg:1}(a) and (b) display the low-temperature
diffraction patterns taken along the [001] and [010] zone-axis
directions, respectively. Both diffraction patterns show notable
superlattice reflections in addition to the  fundamental Bragg
diffraction spots.  Schematic representations of the diffraction patterns,
corresponding to the $a^*$-$b^*$ and $a^*$-$c^*$ planes of reciprocal
space, are shown in Figs.~\ref{fg:1}(c) and (d), respectively.  The
superstructure peaks are characterized by a unique modulation wave vector,
${\bf k} = (2\varepsilon,0,1)$, with $2\varepsilon\approx0.57$,
corresponding to $\varepsilon\approx0.285$, a value consistent with the
neutron work.\cite{lee02}  Note that a more typical [001]-zone-axis
pattern, as shown in Fig.~\ref{fg:2}(a), exhibits a second modulation
wave vector,
${\bf k}' = (0,2\varepsilon,1)$ (see also Ref.~\onlinecite{chen93}).  The
presence of a single modulation wave vector in Fig.~\ref{fg:1}(a) is
clear evidence that the stripe modulation within an NiO$_2$ layer is
one-dimensional.  The occurrence of both ${\bf k}$ and ${\bf k}'$ in
Fig.~\ref{fg:2}(a) is due to the presence of distinct stripe domains
rotated by 90$^\circ$ to one another.  Such twinning is expected, since
the tetragonal lattice provides no unique orientation for the stripe
modulation. 

To further characterize the structural modulation, we have made a
comprehensive examination along several relevant orientations and in many
areas of the sample over the temperature range of 300~K down to 86~K. 
Generally, the  superstructure reflections look very weak and only become
detectable as the temperature is reduced below 200~K, the approximate
charge-ordering temperature.\cite{lee02} The superlattice spots on the
$a^*$-$b^*$ plane are very sharp, indicating a relatively long coherence
length of the ordered state within the NiO$_2$ planes.  The diffraction
pattern along the [010] zone axis exhibits weak diffuse spots that are
extended along the $c^*$ direction, indicating a short correlation length
along the $c$-axis, perpendicular to the NiO$_2$ planes.  The
incommensurability
$\varepsilon$ is observed to vary slightly from one area to another
within the range of 0.28--0.3; similar variations were observed
previously.\cite{chen93} 

Let us return to Fig.~\ref{fg:2}(a) and its orthogonal sets of 
superstructure diffraction spots along the a* and b* directions.  We can
test our assertion that these spots are distinct twin domains by forming
a dark-field TEM image using the circled satellite peak.  If the two sets
of peaks came from the same domain, then we would expect the intensity of
the dark-field image to be uniform.  Instead, the actual image,
Fig.~\ref{fg:2}(b), shows a complex domain-like contrast.  The bright
regions represent domains that contribute to the superlattice peak. 
The dark regions are associated with the orthogonal domains.  From the
size of the bright domains, one can estimate a typical domain size of
$\sim50$~nm.

\begin{figure}[t]
\centerline{\includegraphics[width=3.4in]{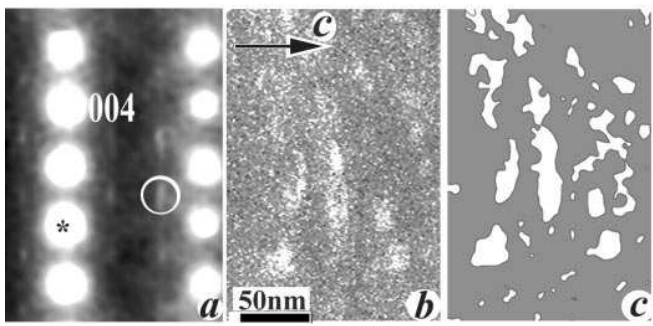}}
\medskip
\caption{(a) [010] zone-axis diffraction pattern showing the diffuse
superlattice reflections in the $a^*$-$c^*$ plane at 86~K. (b) Dark-field
image obtained from the circled satellite spot. The corresponding CO
domains are the bright regions. (c) A smoothed, two-level contour-map
version of (b) to emphasize the charge domains in the examined area. }
\label{fg:3}
\end{figure}

The dark-field images of the charge-ordered state within the $a$-$c$
crystalline plane have also  been checked in many areas at low
temperature.  As mentioned above, the  superlattice spots
generally are weak and diffuse along the direction of the $c^*$-axis as
shown in Fig.~\ref{fg:1}(b). We further illustrate  these features in
Fig.~\ref{fg:3}(a).  The weak, diffuse nature of the spots makes the
measurements of dark-field images at low temperatures quite challenging. 
A very long exposure time is  required to obtain a utilizable dark-field
image, and less-than-perfect mechanical stability of the sample during
such an exposure results in low spatial resolution.  
Figure~\ref{fg:3}(b) shows a dark-field image taken with the 
superlattice reflection circled in Fig.~\ref{fg:3}(a).  Despite the weak
contrast, this image reveals the presence of complex domain structure. 
In order to better depict the structural  properties of the charge
ordered state, we have smoothed the image and created a two-level contour
map, shown in Fig.~\ref{fg:3}(c) illustrating the domain configuration
within the examined  area. This contour map indicates more clearly the
domain features present in Fig.~\ref{fg:3}(b).  As one can see, domains
tend to be elongated along the $a$-axis. Close examination of these
charge order  domains suggests that the longitudinal dimension varies from
10--80~nm. The  transverse dimension, along the $c$-axis, is much
smaller, ranging from 2--5~nm.  The dimensions and orientation of the
domains are consistent with the shape of the
superlattice reflections. 

\begin{figure}
\centerline{\includegraphics[width=3.2in]{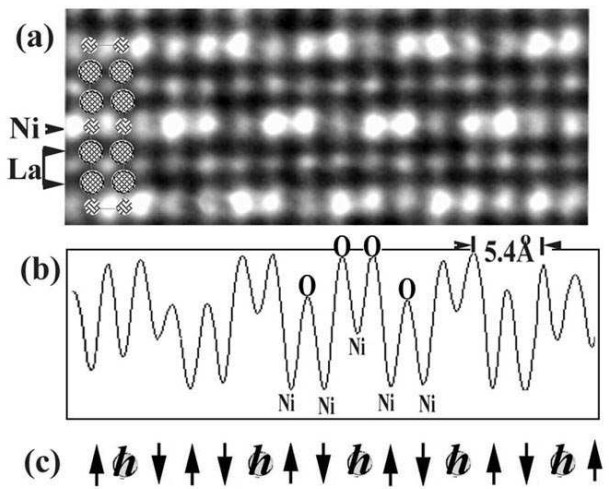}}
\medskip
\caption{(a) High-resolution TEM image of the $a$-$c$ plane ($c$ vertical)
illustrating the modulation at low
temperature. The bright segments arising from charge modulation within
the NiO$_2$ layers can be recognized.  (b) An intensity profile
showing contrast variation within the bottom NiO$_2$ layer of (a).
This curve clearly indicates that the brighter segments in this region are
Ni-centered. (c) Schematic model of spin and charge ordering
corresponding to the curve in (b).}
\label{fg:5}
\end{figure}

\begin{figure}
\centerline{\includegraphics[width=3.2in]{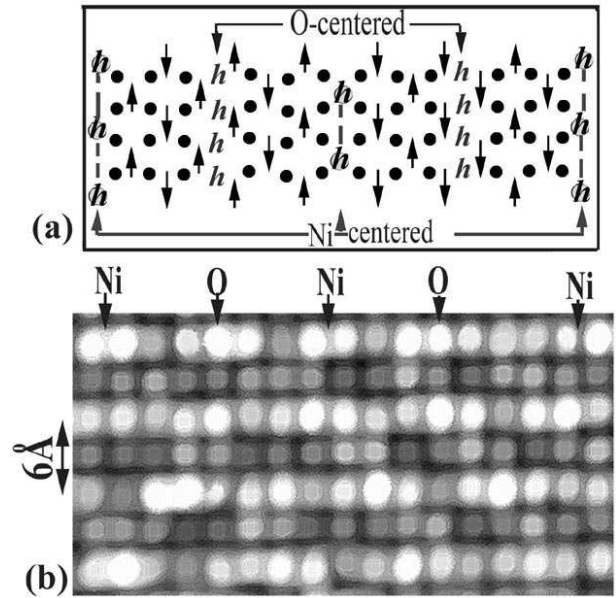}}
\medskip
\caption{(a) Schematic model of alternating Ni- and O-centered stripes
within an NiO$_2$ plane. This configuration corresponds to
$\varepsilon=0.286$.  (Note that the hole density is not properly
represented in the O-centered stripes.)  (b)  High-resolution TEM image of
the $a$-$c$ plane ($c$ vertical) demonstrating the appearance of
alternating O-centered and Ni-centered charge stripes along the
modulation direction. The centers of the charge stripes in the top row
are labelled.  Note that only the first two on the left are aligned with
the charge stripes indicated in (a).}
\label{fg:6}
\end{figure}

\begin{figure}
\centerline{\includegraphics[width=3.2in]{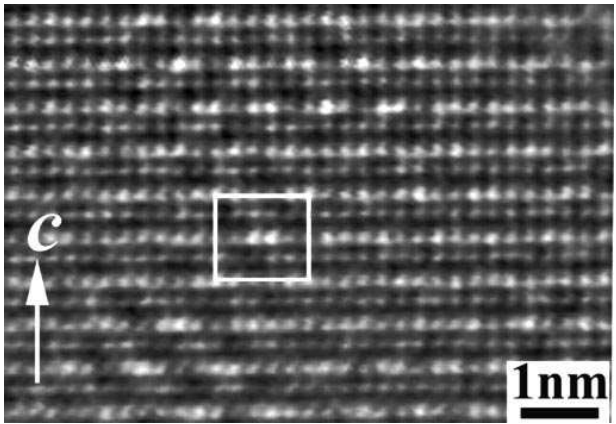}}
\medskip
\caption{High-resolution TEM image showing a larger area in the $a$-$c$
plane. The relative ordering of the charge stripes along the c-axis
direction can be seen.  The white box outlines an approximate unit cell.}
\label{fg:4}
\end{figure}

To study the local ordering of the charge stripes with respect to the
lattice, we turn to high-resolution TEM images.  The images shown in
Figs.~\ref{fg:5}--\ref{fg:4} involve projections along the [010] axis that
were obtained from thin regions of the sample utilizing the Scherzer
defocus condition.  Under these circumstances, the projected rows of
heavier atoms, such as La and Ni, appear as dark spots; their positions
are indicated in Fig.~\ref{fg:5}(a).  Following this identification, it
is clear that the bright spots within each NiO$_2$ plane are associated
with the O rows.  The modulation of the contrast at the O sites should
result from a combination of charge and structural modulations.  To
properly correlate the brighter and darker O spots with hole rich and
poor regions requires a detailed image simulation.  Such a simulation
is challenging because of the need to model both atomic displacements and
charge modulation.  In the absence of a suitable simulation, we simply
note below that associating the brighter O spots with hole-rich columns
yields a satisfying consistency with simple models.  

Figure~\ref{fg:5}(a) shows an image in which the charge stripes, viewed
edge on, appear to be dominantly centered on Ni rows, as indicated by
bright pairs of spots.  The intensity profile along a line through the
bottom NiO$_2$ layer is presented in Fig.~\ref{fg:5}(b).  The Ni rows
correspond to local minima in the contrast, and the O rows to local
maxima.  A corresponding schematic model for the charge and spin stripes
(involving only the Ni sites) is shown in Fig.~\ref{fg:5}(c).

There are also regions of the sample that appear to exhibit alternating
O-centered and Ni-centered stripes; one of these is shown in
Fig.~\ref{fg:6}(b).  We associate O-centered stripes with the occurrence
of three bright spot in a row, with the center one being the brightest. 
Figure~\ref{fg:6}(a) shows a model for equally-spaced, alternating site-
and bond-centered stripes within an NiO$_2$ plane.  Such a configuration
corresponds to $2\varepsilon = 2/3.5$, or $\varepsilon=0.286$, very close
to the average value determined by electron diffraction.  To obtain the
same period with only Ni-centered stripes requires alternating stripe
spacings of $1.5a$ and $2a$, as discussed in Ref.~\onlinecite{woch98}. 
(Note that there are two Ni rows per orthorhombic unit cell.)  Irregular
spacing is apparent in all rows, including the first.

The last issue to be discussed is the stacking of stripes from one layer
to the next.  Figure~\ref{fg:4} shows another image of the $a$-$c$ plane,
but with lower magnification.  Despite plenty of defects, one can observe
many stripes within the layers and their tendency to form a body-centered
stacking along the $c$-axis.  Such an arrangement minimizes the Coulomb
repulsion between the stripes.  The white box outlines an approximate unit
cell.  The body-centering is only approximate because of the pinning of
the stripes to the lattice.  This is especially apparent in
Fig.~\ref{fg:5}(a), where, in moving from one layer to the next, the
Ni-centered stripes can shift only by increments of $a/2$.  For this
sample, with an average stripe spacing of about $1.75a$, it is not
possible to have a perfectly body-centered stacking as long as the
stripes are pinned to the lattice.  This conclusion was first reached in
a neutron diffraction study,\cite{woch98} and the high-resolution TEM
images provide solid confirmation of it.  Finally, we note that the
body-centered stacking makes it impossible to image the stripes directly
in the $a$-$b$ plane, as a projection along the $c$-axis will average out
all stripe contrast.

In summary, we have presented a systematic TEM investigation of
superlattice modulation in \lsnox\ at low temperature.  The observation
of sample regions with a single modulation wave vector provides direct
evidence that the stripe modulation is indeed one-dimensional, and not
grid-like.  This conclusion is also supported by dark-field
images formed from individual superlattice reflections.  A similar
conclusion regarding diagonal spin stripes in La$_{2-x}$Sr$_x$CuO$_4$
with $x\lesssim0.055$ has been obtained by neutron
diffraction\cite{waki00}; in that case, the stripe orientation is tied to
the orthorhombic symmetry of the lattice, which contrasts with the
tetragonal symmetry of our nickelate sample.  Close examination of the
high-resolution TEM images suggests that the charge stripes are
predominantly centered on rows of Ni atoms, although alternative mixtures
of Ni- and O-centered stripes also appear in some small regions.

%
%
Research at Brookhaven is supported by the Department of Energy's (DOE)
Office of Science under Contract No.\ DE-AC02-98CH10886.  KY acknowledges
support from the Japan Science and Technology Corporation, the Core
Research for Evolutional Science and Technology Project (CREST), and
Grants-in-Aid for Scientific Research on Priority Areas, 12046239, 2001
and Research (A), 10304026, 2001 and for Creative Scientific Research
(13NP0201) from the Japanese Ministry of Education, Culture, Sports,
Science, and Technology.  DJB acknowledges support from DOE under
Contract No.\ DE-FG02-00ER45800.


\end{document}